\documentstyle[12pt,worldsci,epsfig]{article}
\newcommand{\mr}{\mathrm}

\newcommand {\oa} {\mbox{${\cal O}(\alpha)$}}

\newcommand{\bq}{\begin{equation}}
\newcommand{\eq}{\end{equation}}
\newcommand{\ba}{\begin{eqnarray}}
\newcommand{\ea}{\end{eqnarray}}
\begin{document}                                                                

\author{\ A.~A.~Akhundov \\
\it DESY - Institut f\"ur Hochenergiephysik Zeuthen,\\
\it Platanenallee 6, D-15738 Zeuthen, Germany \\
 \it and \\
\it Institute of Physics, Azerbaijan Academy of Sciences \\
\it pr.~Azizbekova 33, 370143 Baku, Azerbaijan}
\title{{\bf Radiative Corrections for H1 {\boldmath $F_2(x,Q^2)$} Measurement}}

\begin{center}
{\it Talk given at the Workshop "RADIATIVE CORRECTIONS relevant for 
     the   
  HERMES EXPERIMENT", Zeuthen, Berlin, Germany, 21-23 June, 1993  }
\end{center}

\maketitle                                                                      
\setlength{\baselineskip}{2.6ex}

\begin{abstract}
The numerical study of QED  radiative corrections in the low $x$ region
for the H1 experiment at HERA was performed.
The analytical program  TERAD91 was  used to get
the size of the radiative corrections
for the measurement of the proton structure function $F_2(x,Q^2)$ in the
range  $x= 10^{-2} - 10^{-4}$ and $ 5~GeV^2 <Q^2 < 80~GeV^2$.
It was found that radiative corrections can be not small in some points
of the phase space of the mixed variables.
\end{abstract}

In the long period from the SLAC experiments~\cite{PAN} till the operation of HERA, the
experiments with neutral current deep inelastic $ep$-scattering
%-------------------
\bq
e(k_1) + p(p_1) \rightarrow e(k_2) + X(p_2)
\label{eq0}
\eq
%--------------------
had a relatively simple structure, and
the measurements completely rested on the registration of energy and angle
of the scattered electron.
The new detector generation, the HERA detectors H1~\cite{H1} and ZEUS~\cite{ZEUS},
allows to measure both the scattered electron and hadronic final state.

For the physical analysis of the $ep$-collisions at HERA one can use not only
the familiar electron variables
%-------------
\ba
Q_e^2 = -(k_1 - k_2)^2, \hspace{1.cm} y_e =  p_1 (k_1 - k_2)/ p_1 k_1 ,
\hspace{1.cm} x_e =  Q_e^2/(s y_e),
\label{qxyl}
\ea
%-------------
where
%-------------
\ba
     s = (k_1 + p_1)^2 = 4E_eE_p ,
\label{s}
\ea
%-------------
but also the kinematical variables from the hadron measurement
%-------------
\ba
Q_h^2 = -(p_2 - p_1)^2, \hspace{1.cm} y_h =  p_1 (p_2 - p_1)/ p_1 k_1 ,
\hspace{1.cm} x_h =  Q_h^2/(s y_h),
\label{qxyh}
\ea
%-------------
or some mixture of both. Here $E_e$ and $E_p$ are the energies of incident
electron and proton.

In order to extract from experimental data the inclusive cross section of
deep inelastic $ep$-scattering
in one-boson-exchange approximation one must to take into account contributions
from higher order QED ~\cite{MoTsai} and electroweak processes.
   For example the bremsstrahlung process:
%--------------------
\bq
e(k_1) + p(p_1) \rightarrow e(k_2) + X(p_2) + \gamma(k),
\label{eq00}
\eq
%--------------------
and also the elastic radiative tail :
%--------------------
\bq
e(k_1) + p(p_1) \rightarrow e(k_2) + p(p_2) + \gamma(k).
\label{eqERT}
\eq
%--------------------
contribute to the observed cross section of the reaction (~\ref{eq0}).

For  different choices of the variables  there are differences
in the predictions for the contribution of the higher order processes.

The theoretical investigations of these radiative effects, or radiative corrections (RC),
for experiments at HERA were summarized in Proceedings of the Workshop "Physics at HERA"
 ~\cite{heraproc}. Different programs to calculate the radiative corrections  have been
cross-checked ~\cite{hubert}.

The analysis of neutral current deep inelastic scattering data in the H1 experiment
at HERA in 1992  ~\cite{H1F2}  was based on two independent approaches:

 I) "detector oriented" approach ( "Brussels-Paris-Saclay analysis" )

 II) "physics oriented" approach ( "Zeuthen analysis" )

Within the analysis I the kinematics of the inclusive deep inelastic scattering
process ~(\ref{eq0}) were determined from the energy $E_e'$ of the scattered electron,
and its polar angle
  $\theta_e$,  measured relative to the proton beam direction
%-------------
\ba
Q_e^2   = 4E_eE_e'\cos^2(\theta_e/2),
\label{qe}
\ea
%-------------
%-------------
\ba
y_e =  1 -      (E_e'/E_e)  \sin^2(\theta_e/2).
\label{ye}
\ea
%-------------

The cross sections, acceptances of detector, radiative corrections and resolution
effects were obtained in ($\sqrt{E_e'},\theta_e $) bins.
The cross sections, that have been measured using this binning,
were transformed into cross sections in ($x_e,Q_e^2$)-bins.

In the analysis II the kinematics of the process ~(\ref{eq0}) were obtained from
a measurement of the electron variables
$E_e'$, $\theta_e$ and scaling variable $y_h$ from the hadrons. The hadronic variable
$y_h$  ~(\ref{qxyh}) can be determined using the Jacquet-Blondel method~\cite{JB}
%-------------
\ba
y_h = \sum_{hadrons} \frac{E_h - p_{z,h}}{2 E_e},
\label{yh}
\ea
%-------------
where $E_h$ is the energy of a hadron and $p_{z,h}$
                                                its momentum component
along the incident proton direction.

   The momentum transfer
    $Q^2$ was determined from the electron variables
$\theta_e$ and  $E_e'$ as in ~(\ref{qe}) and Bjorken $x$
 by using a mixture of the electron and hadron measurements ~\cite{Max}
%-------------
\ba
x_m = Q_e^2/(sy_h).
\label{xm}
\ea
%-------------

In the analysis II cross sections, radiative corrections and efficiencies were
calculated in ($x_m,Q_e^2$)-bins.

A further difference between the two methods of the analysis of the data was
the use of two different types of programs --- the Monte Carlo event generators
HERACLES~\cite{HERACLES} in the analysis I and the analytical program 
TERAD91~\cite{TERAD91}
                         in the analysis II --- for the calculations
of the radiative correction factor $ \delta (x,Q^2)$. The radiative correction 
factor $ \delta (x,Q^2)$ is defined by the equation
%-------------------------------------------------------------
\ba
\delta({x,Q^2})=
 \frac{d^{2}{\sigma}^{\mr{theor}}/{dxdQ^2 }}
      {d^{2}{\sigma}^{\mr{Born} } /{dxdQ^2 }}-1 ,
\label{delta}
\ea
%-------------------------------------------------------------
where  $ d^{2}{\sigma}^{\mr{theor}}/{dxdQ^2} $ is the
theoretical approximation to the measured cross section
~$ d^{2}{\sigma}^{\mr{meas}}/{dxdQ^2} $ , which contains the contributions from
higher order electroweak processes, and
    $ d^{2}{\sigma}^{\mr{Born}}/{dxdQ^2} $
is the Born cross section of the process
(\ref{eq0}).

In this note we present  the numerical calculations
                              of the RC factor $\delta (x,Q^2)$
for the "Zeuthen analysis" of the proton structure function $F_2(x,Q^2)$ in the
range  $x= 10^{-2} - 10^{-4}$ and $5~GeV^2 < Q^2 < 80~GeV^2$ in the H1 experiment
  at HERA   ~\cite{H1F2}.

 The Monte Carlo HERACLES is suited for the calculation of RC  to the differential
 cross sections in terms of the electron variables because the program calculates cross sections
 and generates events in terms of the  variables $x_e$ and $Q^{2}_{e}$.
 Therefore most of the generated radiative events in HERACLES have low $Q_h^2$ and
 the calculation of the radiative corrections in the terms of other 
 variables ( hadron, mixed ) can be performed
 with less efficiency only.
 For this reason in the analysis II was used  the  analytical program TERAD91.

      TERAD91   - a program package to calculate radiative corrections for
deep inelastic $ep$-scattering - was created in 1991 during Workshop "Physics at HERA"
   and accumulates about 15 years experience in the field
of calculation of the radiative corrections for deep inelastic $lN$-scattering
 ~\cite{exper}.
 The program   TERAD91  can be applied to calculate the RC factor of the order {\oa}
to the measured differential cross section,
                            $d^2 \sigma ^{meas} /dxdQ^2$,
after the acceptance and smearing corrections.

 The program package TERAD91 contains one part, TERAD,
   which is based on
   the model-independent treatment ~\cite{MI} of the leptonic QED corrections
            for neutral current $ep$-scattering,
 including bremsstrahlung process (~\ref{eq00}) and QED vertex corrections.
 TERAD does not necessarily rely on the quark parton model and uses the
 phenomenological structure functions $F_1, F_2$ and $F_3$, which are defined in the
 Born approximation.

          The program TERAD91 can be used  for the model-independent calculations
of QED  corrections for neutral current deep inelastic $ep$-scattering 
  in the terms of several variables - electron, hadron, mixed variables ~\cite{teup}-
 and in the new version TERAD93 ~\cite{TERAD93} also
                                             in terms of the Jacquet-Blondel variables ~\cite{jbpl}.
The leptonic QED corrections are dominated  and must be calculated
carefully.

In the low $Q^2$ region the Born cross section of the deep inelastic scattering (\ref{eq0})
is determined by two structure function  $F_2$ and $2xF_1=F_2/(1+R)$:
%----------------------------------------------------
\ba
 \frac{d^2\sigma}{dxdQ^2} = \frac{2\pi \alpha^2}{Q^4 x}
\left( 2~(1-y)~+~\frac{y^2}{1+R} \right)   F_2(x,Q^2) .
\label{Born}
\ea
%-------------------------------------------------------------

 In order to calculate the RC one has to use a realistic parametrization of the
 structure functions $F_i(x,Q^2)$ over the full range of $x$ and $Q^2$.
 We have used several different parameterizations of the deep inelastic proton
                                           structure functions ,
 including MRS D- and MRS D0 parameterizations ~\cite{MRSD}.
 For region $Q^2<5~GeV^2$ the parameterizations were taken at $Q^2=5~GeV^2$ and multiplied by
                                    factor~\cite{Prokh}
 $ (1 - exp(-aQ^2)) $ , $ a=3.37~GeV^{-2}$. In our calculations we have assumed $R=0$.

 In the "Zeuthen analysis"  the following $x,Q^2$-binning was used.
 There were four $Q^2$-bins:
  \ba
       Q^2 = 5 - 10, 10 - 20, 20 - 40, 40 - 80~ GeV^2
  \label{binq}
  \ea
    and four bins per $x$-decade:
  \ba
       log~x  = -4.0, -3.75, -3.50, -3.25, ..., -1.0 .
  \label{binx}
  \ea

 The kinematical region of the experimental measurement was defined by:
  \ba
      157.5^o < \theta_e < 172.5^o, \hspace{1.cm}
       0.05 < y_e < 0.60
  \label{theye}
  \ea
 for electron variables and
  \ba
      157.5 < \theta_e < 172.5^o, \hspace{1.cm}
              y_h < 0.50 ,  \hspace{1.cm}
              \theta_{jet} > 10^o
  \label{theyh}
  \ea
 for mixed variables.

 Figs.1,2   show the results for the radiative corrections in the terms of the mixed variables
 $x_m,Q_e^2$.
which are calculated by the integrations over bins.
 The RC in the terms of the mixed variables,
 $ \delta  (x_m,Q^2_e)$,  are not small   and can be reach 7 \%
 at $x=10^{-2}$.

  The "Zeuthen analysis"  of the deep inelastic scattering has included also the analysis
                                                         of the experimental data in the terms
  of the electron variables $x_e,Q_e^2$ corresponding to the binning (~\ref{binq}),(~\ref{binx}).
  Fig.3,4 show the results of the calculations of the RC in the bins $x_e,Q_e^2$, that  have been
  obtained with the program TERAD91.

    The corrections     determined by the
  electron variables, $\delta (x_e,Q^2_e)$,
     are known to be large at low $x$   and dominated by hard
  photon emission from the incoming electron. From the comparison of Fig.3 and Fig.4
  it is seen that the sensitivity of the RC to the choice of the structure functions
 is not small.

The comparison of the results for the radiative corrections in the terms of the
electron variables $E_e',\theta_e$,   that have been obtained ~\cite{note} with
event generator HERACLES ( interface DJANGO~\cite{DJANGO}) and analytical program TERAD93,
is shown on Fig.5  for the bin
                           $162.5^{o} < \theta_{e} < 167.5^{o}$.
Here the variable  $u$ is defined  by the equation
\ba
u = 40 \sqrt{ E_e'/E_e},
\label{u}
\ea
 and for the structure function $F_2$ the parametrization MRS D- was used and $R$=0.
The results of the both programs are in good agreement.

Now we will discuss the influence of the calorimetric measurement of the energy
of the scattered electron  on the size of the radiative corrections.

Because the collinear radiation of photons
in the final state of the reaction (~\ref{eq00}) is not resolved in the electromagnetic
calorimeter, one must calculate the RC in terms of the calorimetric variables
including the total energy of the electromagnetic cluster
  \ba
      E_{vis} = E_e' + E_{\gamma}.
  \label{vis }
  \ea

 Fig.6 shows the results for the radiative corrections including the effect of the
 calorimetric measurement of the final state photon radiation obtained ~\cite{note} with
 the Monte Carlo HERACLES. For comparison the RC   in the terms
 of the pure electron variables are shown also.

 In the program TERAD93 the possibility  to calculate the RC
 in the terms of calorimetric variables, $ E_{vis},\theta_e $, has not been included yet,
    but with the help of the leading log calculations ~\cite{Blum}
                   one can estimate the size of the influence of
      the calorimetric measurement on the radiative corrections.

 The calorimetric measurement cannot distinguish between a single electron and
 an electron with accompanying photons. Therefore the final state radiation has
 experimentally no effect on the measurement of the kinematics of deep inelastic
 scattering    and can be absorbed in the definition of the final
 electron energy of the process (~\ref{eq0}).

 This leads to the following way ~\cite{note} to take into account the calorimetric measurement.
 We can calculate  the contribution of the final state radiation in the leading
 logarithmic approximation  and subtract from the total RC, that we
 obtain with the program TERAD93.

    The more detailed investigation of the calorimetry
 problem will be present in the separate publication.

\section*{Acknowledgements}
 
I  would like to thank D.~Bardin and M.~Klein for numerous helpful discussion.
It is a pleasure to thank the DESY--IfH~Zeuthen for the good
opportunities to work at DESY.

%
%%%%%%%%%%%%%%%%%  references %%%%%%%%%%%%%%%%%%%%%%%%%%%%%%%%%%%%%%%%%%
%

\begin{figure}[htbp]
\begin{picture}(160,180)
\put(-35,-90){\epsfig{file=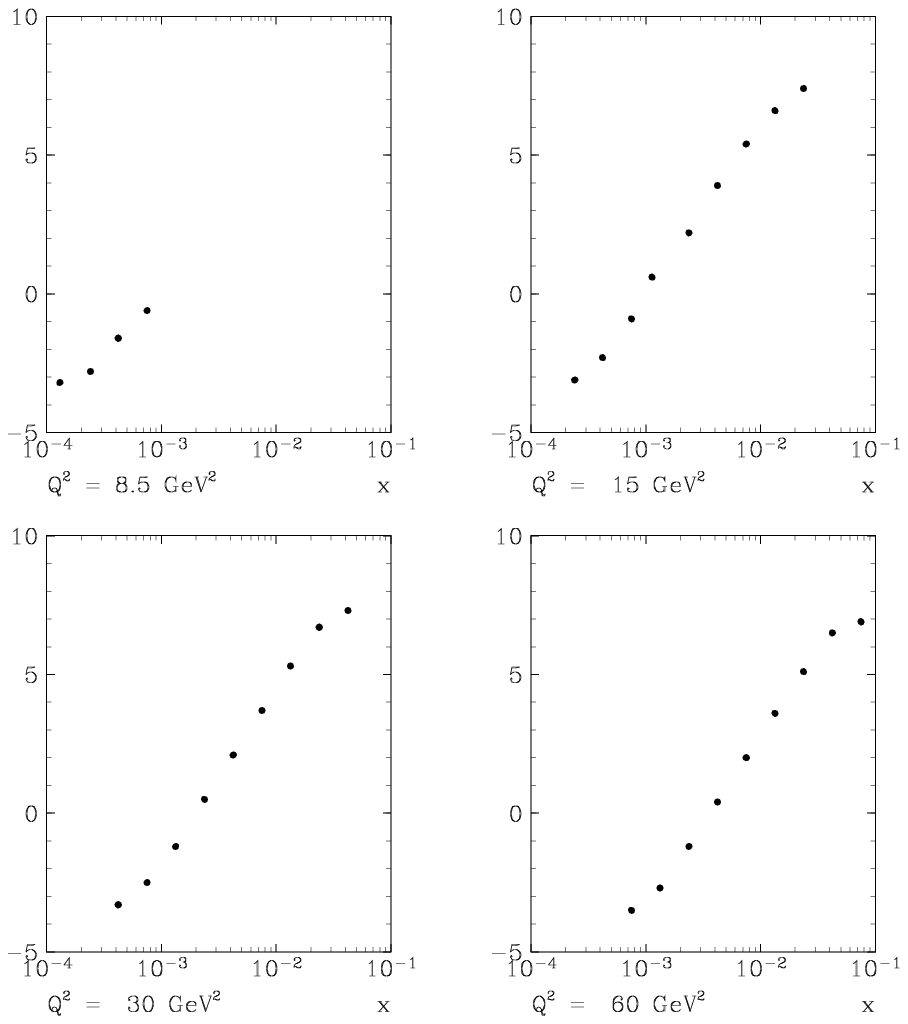,width=200mm}}
\end{picture}                                                                   
                                                                                
\caption[]{ Radiative correction factor
 $ \delta  (x_m,Q^2_e)$ (\%)in $(x,Q^2)$-bins for MRS D- parametrization.}
\end{figure}

\begin{figure}[htbp]
\begin{picture}(160,180)
\put(-35,-90){\epsfig{file=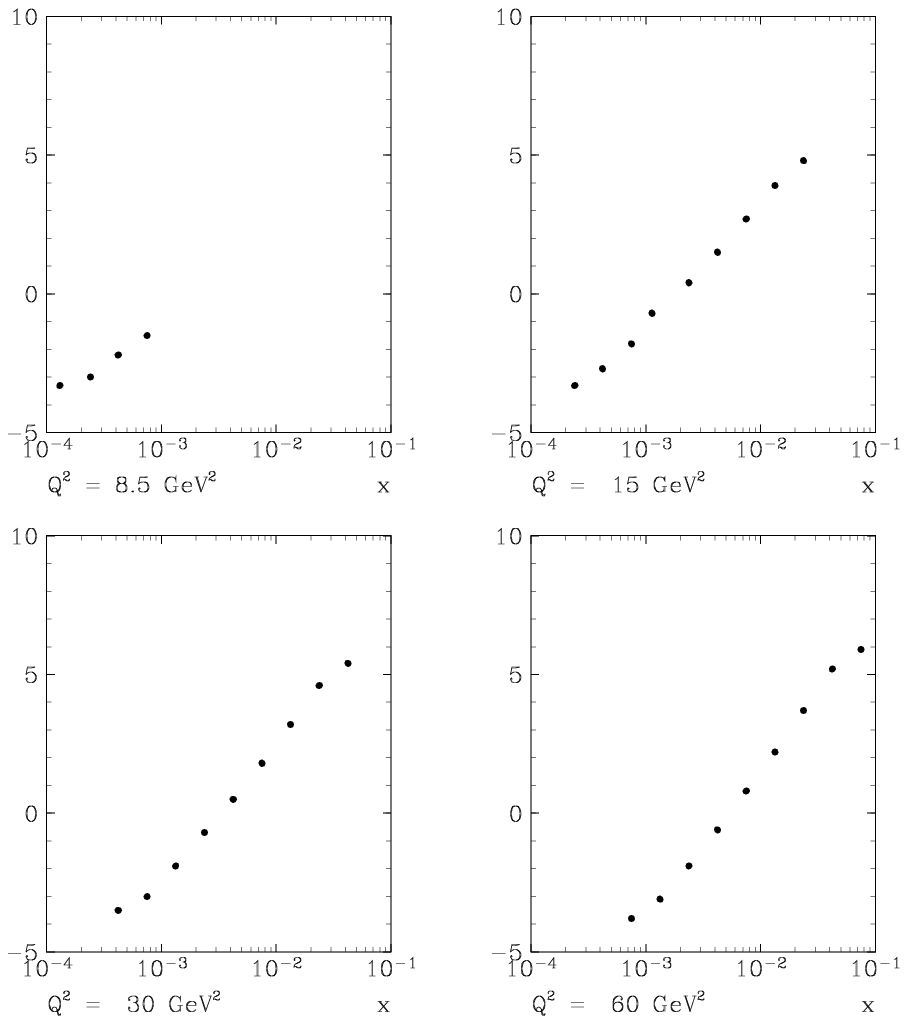,width=200mm}}
\end{picture}                                                                   
                                                                                
\caption[]{ Radiative correction factor
 $ \delta  (x_m,Q^2_e)$ (\%)in $(x,Q^2)$-bins for MRS D0 parametrization.}
\end{figure}

\begin{figure}[htbp]
\begin{picture}(160,180)
\put(-35,-90){\epsfig{file=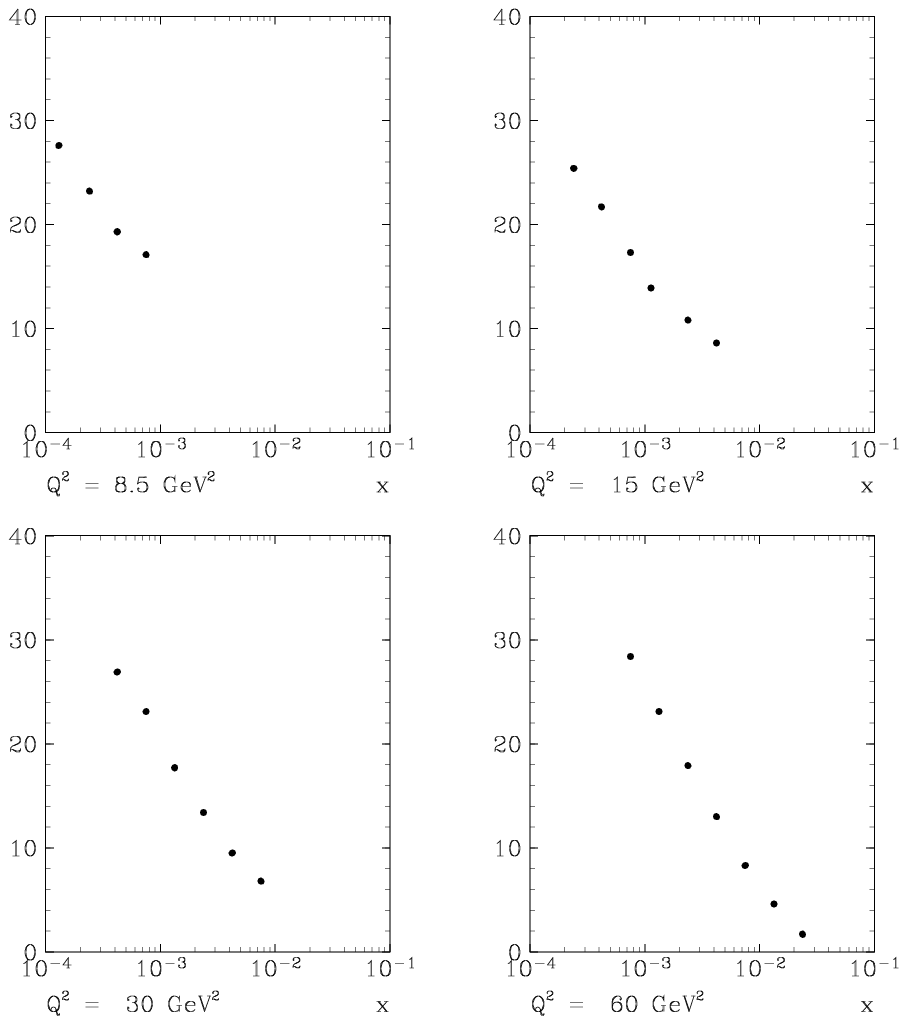,width=200mm}}
\end{picture}                                                                   
                                                                                
\caption[]{ Radiative correction factor
 $ \delta  (x_e,Q^2_e)$ (\%)in $(x,Q^2)$-bins for MRS D- parametrization.}
\end{figure}

\begin{figure}[htbp]
\begin{picture}(160,180)
\put(-35,-90){\epsfig{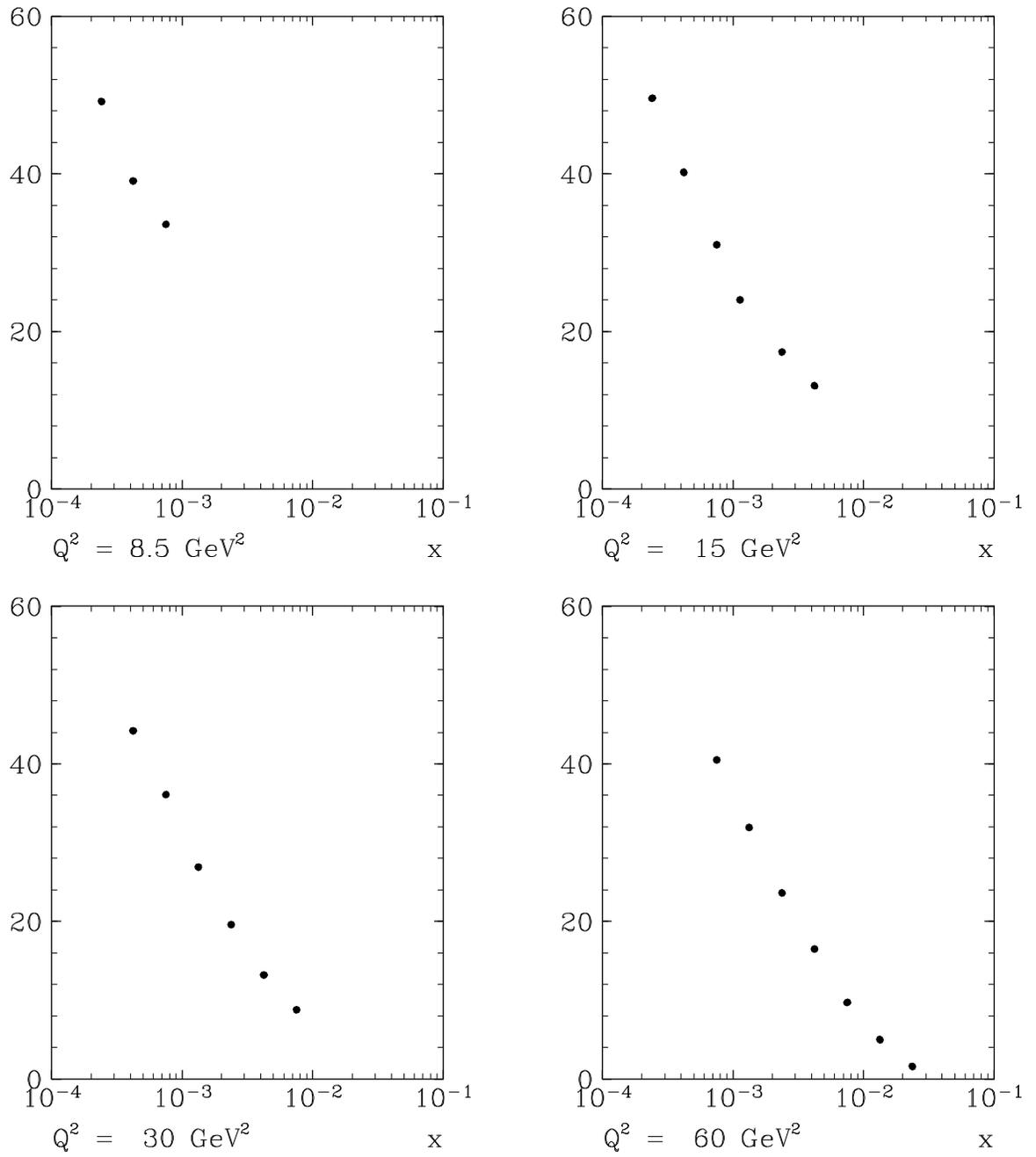}}
\end{picture}                                                                   
                                                                                
\caption[]{ Radiative correction factor
 $ \delta  (x_e,Q^2_e)$ (\%)in $(x,Q^2)$-bins for MRS D0 parametrization.}
\end{figure}

\begin{figure}[htbp]
\begin{picture}(160,180)
\put(-5,-15){\epsfig{file=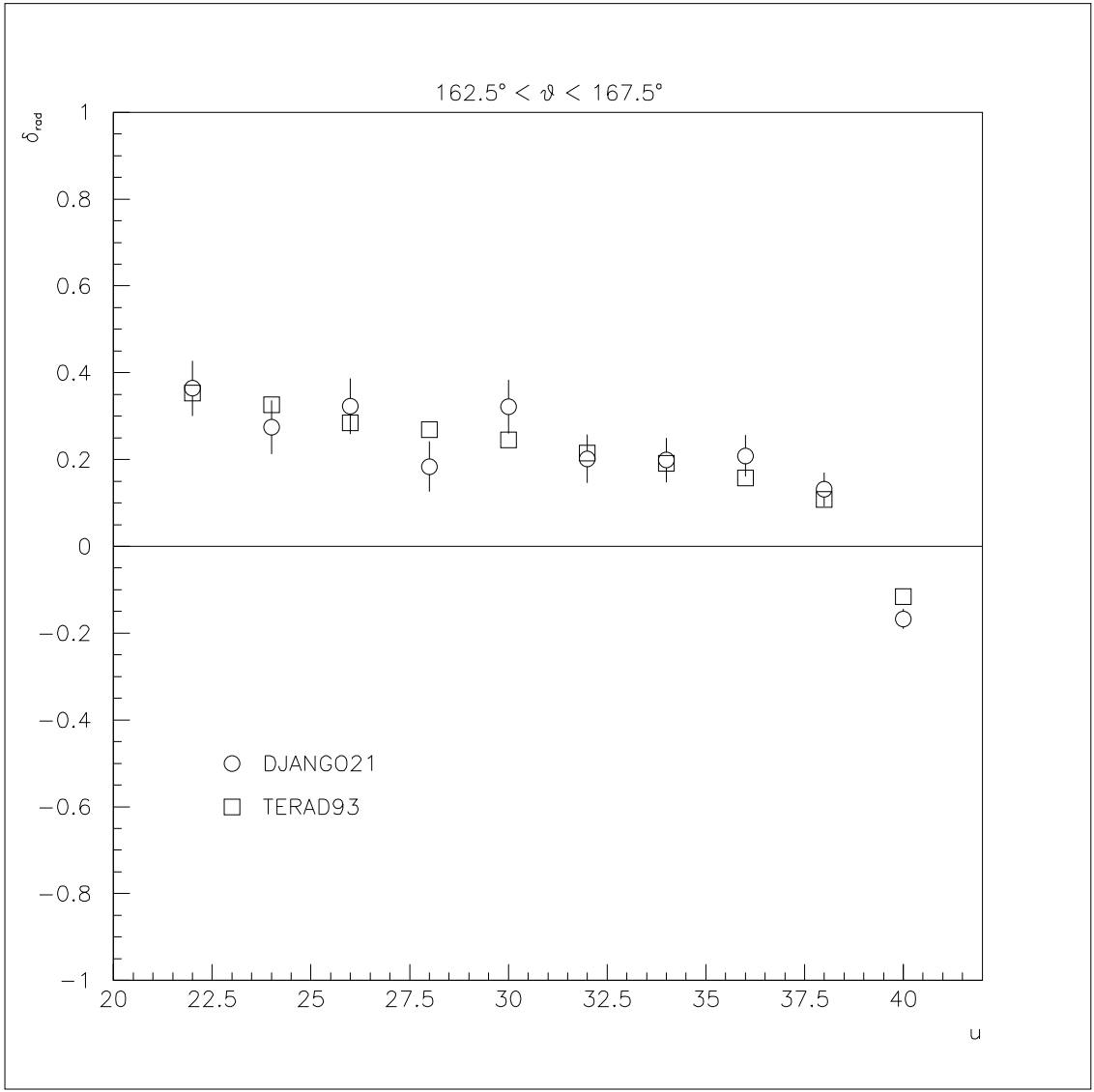,width=150mm}}
\end{picture}                                                                   

\caption[]{  Radiative correction
 $ \delta  (u,\theta_e)$ in $ u $-bins for
                           $162.5^{o} < \theta_{e} < 167.5^{o}$.}
\end{figure}
                                                                                
\begin{figure}[htbp]
\begin{picture}(160,180)
\put(-5,-15){\epsfig{file=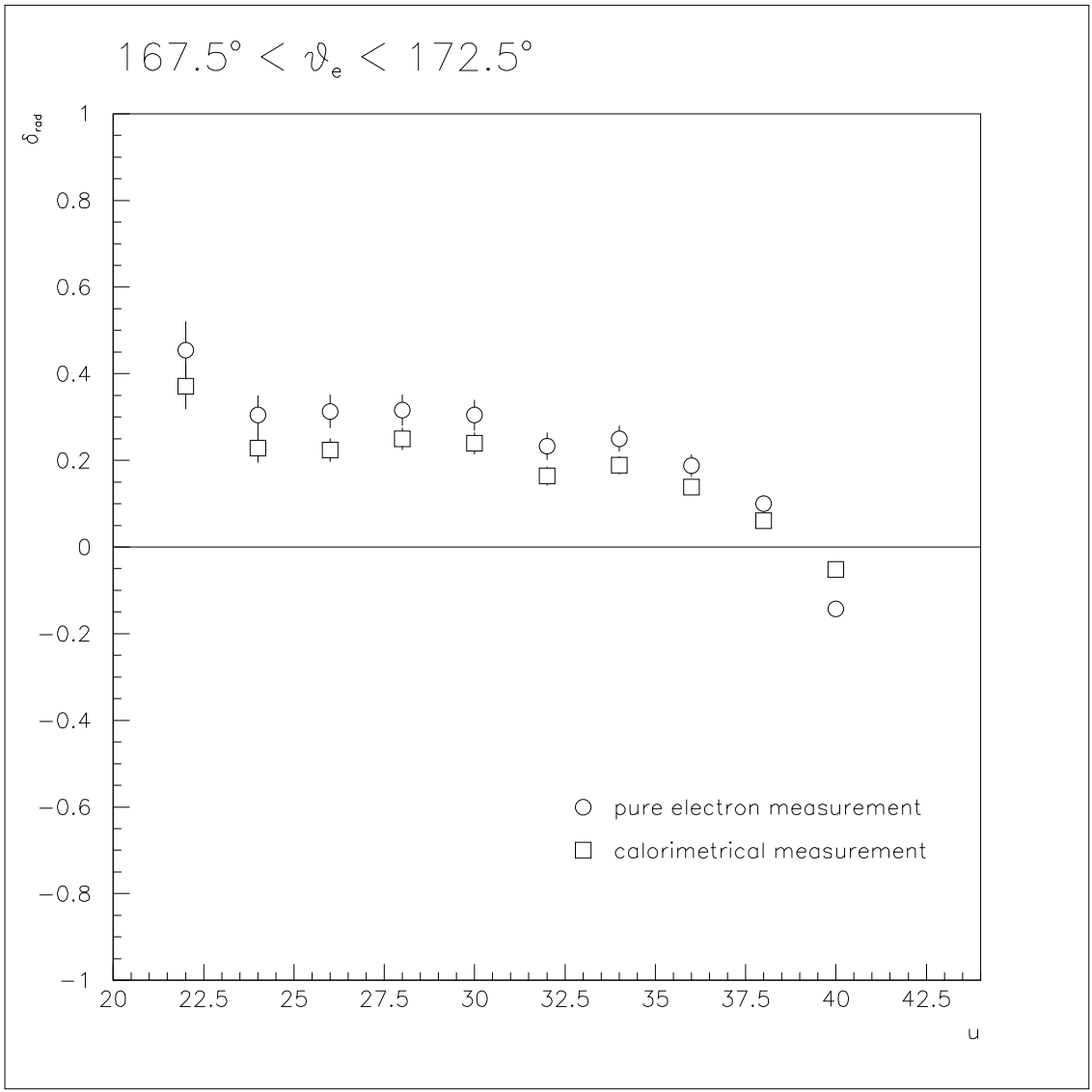,width=150mm}}
\end{picture}                                                                   

\caption[]{ Effect of calorimetric measurement on size of radiative
 correction $ \delta  (u,\theta_e)$ \\
                      for  $167.5^{o} < \theta_{e} < 172.5^{o}$.}
\end{figure}
                                                                                
\end{document}